\documentclass[journal]{IEEEtran}
\usepackage{mathrsfs}
\usepackage{url,times,amsmath,color,amssymb,graphicx,epsfig,psfig,cite,psfrag,subfigure,dsfont,booktabs}
\usepackage{algorithm}
\usepackage{algorithmic}
\usepackage{bm}
\usepackage{stfloats}
\usepackage{multirow}

\begin{document}
\title{Deep Learning based Radio Resource Management in NOMA Networks: User Association, Subchannel and Power Allocation}





\author{Haijun Zhang,~\IEEEmembership{Senior Member,~IEEE}, Haisen Zhang,\\ Keping Long,~\IEEEmembership{Senior Member,~IEEE} and George K. Karagiannidis,~\IEEEmembership{Fellow,~IEEE}

\thanks{Haijun Zhang, Haisen Zhang and Keping Long are with the Institute of Artificial Intelligence, Beijing Advanced Innovation Center for Materials Genome Engineering, Beijing Engineering and Technology Research Center for Convergence Networks and Ubiquitous Services, University of Science and Technology Beijing, Beijing, 100083, China (e-mail: haijunzhang@ieee.org, longkeping@ustb.edu.cn).


George K. Karagiannidis is with Aristotle University of Thessaloniki, Thessaloniki, Creece (e-mail: geokarag@ieee.org).


}} \maketitle

\pagestyle{empty}  
\thispagestyle{empty} 

\begin{abstract}
With the rapid development of future wireless communication, the combination of NOMA technology and millimeter-wave(mmWave) technology has become a research hotspot.
The application of NOMA in mmWave heterogeneous networks can meet the diverse needs of users in different applications and scenarios in future communications.
In this paper, we propose a machine learning framework to deal with the user association, subchannel and power allocation problems in such a complex scenario.
We focus on maximizing the energy efficiency (EE) of the system under the constraints of quality of service (QoS), interference limitation, and power limitation.
Specifically, user association is solved through the Lagrange dual decomposition method, while semi-supervised learning and deep neural network (DNN) are used for the subchannel and power allocation, respectively.
In particular, unlabeled samples are introduced to improve approximation and generalization ability for subchannel allocation.
The simulation indicates that the proposed scheme can achieve higher EE with lower complexity.
\end{abstract}

\section{Introduction}
With the explosively growth of users' service requirements, the use of the limited radio resources to hold more wireless service of next generation wireless networks becomes a challenging problem.
Non-orthogonal multiple access (NOMA) is a novel access technique for future communication system, which has been widely concerned because of its high spectral efficiency.
The application of NOMA in heterogeneous
networks can meet the diverse needs of users in future communications.
Resource management in NOMA networks is becoming more and more intriguing to enhance energy efficiency (EE) of systems.
Existing research studied several resource management schemes for NOMA networks.
Most of the existing method requires high computation and may not be not necessarily feasible in practice.

In existing research, dual decomposition method was widely used to user association in heterogeneous networks \cite{UA2013,JSAC2017}.
In \cite{UA2015}, an optimization scheme for user association, based on graph theory, was proposed.
Furthermore, a theoretical mean proportionally fair utility was suggested in \cite{UA2017} to address the problem of user association.
In the subchannel allocation problem of NOMA network, existing works often used matching theory.
Especially, a two side many-to-many matching algorithm was proposed in \cite{Noma,Noma2,NomaJ,Noma3}, static and dynamic source-destination  matching algorithm was proposed in \cite{NomaS}.
For the problem of power optimization in NOMA heterogeneous network, there are several power optimization algorithms proposed, such as water-filling algorithm\cite{WPC1}, Lagrange dual multiplier method\cite{PC1}, alternating direction method of multipliers (ADMM) algorithm\cite{ADMM2015}, Lyapunov Optimization algorithm\cite{LPC1}\cite{LPC2}.
However, these methods are often based on high complexity and need many iterations to converge for BS processing.

As the most promising technology in artificial intelligence, machine learning (ML) could be used to address wireless resource management in wireless network \cite{ML2017}.
In this case, the challenge is how to enable ML to assist wireless devices in intelligent learning and decision-making, so as to meet the diverse service demands of the future wireless networks \cite{DL2018}\cite{DL2019}.

Recently, as the representative technology of ML, supervised learning has been used to wireless communications.
Full-connected deep neural network (DNN) is the most classical model of supervised learning. Its non-linear approximation performance enables DNN scheme to solve many problems in resource management, for instance, beamforming, subchannel allocation and power control.
The work in \cite{haoransun2018} was the classic application of DNN scheme in resource management, which used deep learning to approximate the transmit power policy, while it was shown the universal approximation for power optimization.
In \cite{HengtaoHe2018}, a deep learning algorithm in cognitive radio networks was proposed, by considering EE, spectrum efficiency and computing efficiency.
The team DNN was proposed in \cite{TDNN}, which can solve team decision problems of power control and provides a basis for the data parallel distributed neural network in resource management.
In \cite{CNN2018}, a convolutional neural network was proposed for power control, which can achieve higher EE and spectrum efficiency in a much lower computation time,  due to sharing the convolution kernel parameters and the sparsity of inter-layer connections in hidden layers.

As shown in the previous study \cite{haoransun2018,HengtaoHe2018,TDNN,CNN2018}, DNN can be used without explicitly solving the problem of complex optimal control strategy of communication system.
As an intelligent resource management tool, deep learning is used to solve various problems, such as user connection, wireless access selection, frequency allocation, power control and intelligent beamforming.
In addition, considering that the training DNN model with medium number of layers can be used in lower calculation time \cite{Ye}, it is very suitable for real-time operation.
Compared with the conventional distributed optimization technology, the resource management algorithm based on deep learning can understand the wireless network state and network user state in real time, so as to adjust the resource management scheme in real time. This kind of intelligent decision-making is very important for most Internet of things and beyond 5th generation (B5G) services, especially those that need real-time, low latency operation, such as automatic driving and UAV control \cite{C1,C2,ad,vn}.

Nowadays, there are some literatures on DL-based resource management in noma network.
The research of \cite{rv2} used  deep recurrent neural network to solve the resource allocation for the NOMA heterogeneous IoT network.
However, the authors did not consider the influence of high power consumption on EE.
The work in \cite{rv3} proposed the power allocation scheme for NOMA system via exhaustive search method and DNN model.
However, it is unrealistic to obtain training sets by exhaustive methods when the network environment becomes more complex.
The research of \cite{rv4} proposed a deep reinforcement learning framework to allocate channel in a near optimal way.
However, the size of state space will limit the performance of the algorithm in this work.

In our work, we study the problem of wireless resource management aiming at maximizing EE considering power consumption, and the data set obtained by gradient iteration algorithm has advantages in complexity.

The effect of deep learning algorithm largely depends on the label samples. This will inevitably be affected by the process of existing algorithms.
In resource management based on ML, it is simple to obtain massive unlabeled samples, but in this case we need to use expensive and very long computation to get the output of each sample.
Therefore, we try to improve the learning performance and reduce the dependence on the reference algorithm. Thus, semi-supervised learning is introduced to study this optimization problem, that is, training through a finite number of labeled samples generated by numerical iterative algorithms, and these labeled samples are also combined with massive unlabeled samples. Semi-supervised learning can avoid the waste of data and resources  and solve the problem of weak generalization ability of supervised learning model.

Co-training is proposed in \cite{CO} as a methods of semi-supervised learning, which initializes more than two learners, based on clustering or popular hypothesis. In the process of learning, the labeled data with the highest confidence after labeling is selected, and the labeled data is put into each other  after labeling in order to update the model.

According to the author' knowledge, the issue of deep learning based user association, subchannel and power allocation has not been yet well investigated in NOMA networks.
The main contributions of this paper can be summarized as follow:
\begin{itemize}
  \item A deep learning-based framework is proposed, which is mainly used to deal with energy-efficient user association, subchannel and power allocation in NOMA mmWave heterogeneous networks.
  \item Under the premise of guaranteeing quality of service (QoS), interference limitation and power limitation, we concentrate on maximizing the EE of the network. Specifically, sample data are generated by iterative algorithm, and ML schemes are adopted in the decision-making stage.
  \item The Lagrange dual decomposition method based user association scheme is proposed, while semi-supervised learning and DNN are used to address the subchannel allocation and power control in NOMA heterogeneous networks.
  \item Extensive simulations reveal that by using unlabeled data, the proposed method can optimize EE and reduce the dependence on the reference algorithm.
\end{itemize}

The rest of the paper is organized as follows. The system model is provided in Section II.
Section II provides the system model. The deep learning for wireless resource management is proposed in Section III, a large number of simulation experiments were carried out in Section IV,  and the performance of the proposed algorithm was evaluated. Section V summarizes the paper.

\section{System Model and Problem Formulation}

\subsection{System Model}

\begin{figure}[t]
        \centering
        \includegraphics*[width=8cm]{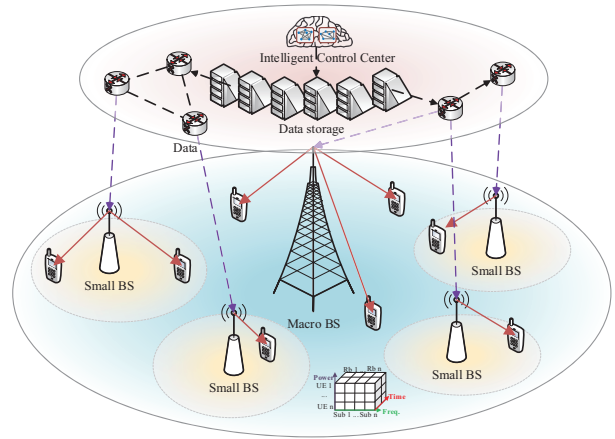}\\
       \caption{A NOMA heterogeneous network with an intelligent control center.}
        \label{fig:1222}
\end{figure}

In this paper, a NOMA based mmWave heterogeneous wireless networks with an intelligent control center (Fig. \ref{fig:1222}) is studied, where small cells are distributed uniformly in a macrocell. We concentrate on issues of downlink user association, subchannel allocation and power control.
Denote $\mathcal{B} = \left\{ {1,2,...,b,...,B} \right\}$ as  the set of all BSs including macro BS and small BSs, $\mathcal{N} = \left\{ {1,2,...,N} \right\}$ is the set of all subchannel, $\mathcal{M} = \left\{ {1,2,...,m,...,M} \right\}$ is the set of users,
and $X = \left\{ {{x_{1,1}},{x_{1,2}}, \cdots ,{x_{b,m}}} \right\}$ is used to represent the user association strategy between BS $b$ and user $m$. If user $m$ is assigned to the BS $b$, then, ${x_{b,m}} = 1$. Otherwise, ${x_{b,m}} = 0$.

It is assumed that a subchannel in NOMA can be utilized by up to two users through successive interference cancelation (SIC).
The set $S = \left\{ { {s_{1,1}^1} ,{s_{1,2}^1}, \cdots ,{s_{b,m}^n}} \right\}$ stands for the subchannel allocation strategy between user $m$ and subchannel $n$ of BS $b$.
If the subchannel $n$ of BS $b$ is occupied by the user $m$, then, ${s_{b,m}^n} = 1$. Otherwise, ${s_{b,m}^n} = 0$.

The signal to interference plus noise ratio (SINR) of user $m$ on subchannel $n$ of BS $b$ can be given as
\begin{equation}
\gamma_{b,m}^n = \dfrac{{s_{b,m}^np_{b,m}^ng_{b,m}^n}}{{g_{b,m}^n\sum\limits_{r = m + 1}^{ M} {s_{b,r}^n p_{b,r}^n}  + \sum\limits_{j = 1,j \ne b}^{ B} {\sum\limits_{r = 1}^{ M} {s_{j,r}^np_{j,r}^n} } g_{j,m}^n + {\sigma ^2}}},
\end{equation}
where ${p_{b,m}^n}$ is the power of user $m$ on subchannel $n$ of BS $b$, and $g_{b,m}^n$ is the gain of user $m$ on subchannel $n$ of BS $b$ and ${\sigma ^2}$ is the variance of AWGN.
The capacity of user $m$ on subchannel $n$ of BS $b$ can be expressed as
\begin{equation}
c_{b,m}^n = \dfrac{W}{{\sum\limits_{m \in {\cal M}} {{x_{b,m}}} }}{\log _2}\left( {1 + \gamma _{b,m}^n} \right),\forall b \in {\cal B},
\end{equation}
where $W$ is the system bandwidth.

The total rate of the network is
\begin{equation}
R\left( {X,S,P} \right) = \sum\limits_{b = 1}^{ B} {\sum\limits_{m = 1}^{ M} {\sum\limits_{n = 1}^{ N} {{x_{b,m}}c_{b,m}^n} } },
\end{equation}
and the total transmit power is
\begin{equation}
U\left( {X,S,P} \right){\rm{ = }}\sum\limits_{b = 1}^{ B} {\sum\limits_{m = 1}^{ M} {\sum\limits_{n = 1}^{ N} {{x_{b,m}}s_{b,m}^np_{b,m}^n{\rm{ + }}} } } \sum\limits_{b = 1}^{ B} {{p_{c,b}}},
\end{equation}
where ${p_{c,b}}$ denotes the circuit power of BS $b$.

The total EE of the system can be formulated as
\begin{equation}
EE = \dfrac{{R\left( {X,S,P} \right)}}{{U\left( {X,S,P} \right)}}.\
\end{equation}

\subsection{Problem Formulation}

It's assumed that each user can be serviced only by one BS at time. Thus,
\begin{equation}
\sum\limits_{b \in {\mathcal{B}}} {{x_{b,m}}}  = 1,\forall m \in \mathcal{M}.
\end{equation}

This work also assumes that at most two users can occupy one subchannel of each BS. Therefore hold that
\begin{equation}
\sum\limits_{m \in \mathcal{M}} {s_{b,m}^n \le 2,} \forall n \in \mathcal{N},b \in \mathcal{B},
\end{equation}
with the constraint:
\begin{equation}
\sum\limits_{m \in {\mathcal{M}}} {{x_{b,m}}}  \le K_b,\forall b \in \mathcal{B},
\end{equation}
where $K_b$ is the the maximum number of users assigned to BS $b$.
The total power constraint is:
\begin{equation}
{\sum\limits_{m = 1}^{ M} {\sum\limits_{n = 1}^{ N} {{x_{b,m}}s_{b,m}^np_{b,m}^n} } }\le {p_{\max }},\forall b \in \mathcal{B},
\end{equation}
where ${p_{\max }}$ is the maximum transmit power of BS $b$,
and the QoS constraint is:
\begin{equation}
{\sum\limits_{m = 1}^{\cal B} {\sum\limits_{n = 1}^{\cal N} {{x_{b,m}}s_{b,m}^nc_{b,m}^n} } }\ge {R_{t}},\forall m \in \mathcal{M},
\end{equation}
which ensures that each user has the minimum QoS requirement of ${R_t}$.
Finally, the cross-tier interference constraint can be written as:
\begin{equation}
\sum\limits_{j = 1,j \ne b}^{ B} {\sum\limits_{r = 1}^{ M} {s_{j,r}^np_{j,r}^n} } g_{j,m}^n \le {I_b},\forall b \in \mathcal{B},
\end{equation}
where ${I_b}$ denotes the maximum interference constraint.

Aiming to achieve load-balance and maximum EE, the problem of user association, subchannel and power optimization can be formulated as:
\begin{equation}
\mathop {\max }\limits_{\left( {X,S,P} \right)} \sum\limits_{b = 1}^B {\sum\limits_{m = 1}^M {\sum\limits_{n = 1}^N {\dfrac{{{x_{b,m}}\log \left( {\dfrac{W}{{\sum\limits_{m \in M} {{x_{b,m}}} }}{{\log }_2}\left( {1 + \gamma _{b,m}^n} \right)} \right)}}{{\sum\limits_{b = 1}^B {{p_{c,b}}}  + \sum\limits_{b = 1}^B {\sum\limits_{m = 1}^M {\sum\limits_{n = 1}^N {{x_{b,m}}s_{b,m}^np_{b,m}^n} } } }}} } }
\label{EE}
\end{equation}
\begin{equation}
\begin{aligned}
& \text{s.t.}
& & {C1:\sum\limits_{b \in {\mathcal{B}}} {{x_{b,m}}}  = 1,\forall m \in \mathcal{M} },\\
&&& {C2:\sum\limits_{m \in \mathcal{M}} {s_{b,m}^n \le 2,} \forall n \in \mathcal{N},b \in \mathcal{B}},\\
&&& {C3:\sum\limits_{m \in {\mathcal{M}}} {{x_{b,m}}}  = K_b,\forall b \in \mathcal{B}},\\
&&& {C4:{\sum\limits_{m = 1}^{ M} {\sum\limits_{n = 1}^{ N} {{x_{b,m}}s_{b,m}^np_{b,m}^n} } }\le {p_{\max }},\forall b \in \mathcal{B}},\\
&&& {C5:{\sum\limits_{m = 1}^{ B} {\sum\limits_{n = 1}^{ N} {{x_{b,m}}s_{b,m}^nc_{b,m}^n} } }\ge {R_{t}},\forall m \in \mathcal{M}},\\
&&& {C6:\sum\limits_{j = 1,j \ne b}^{ B} {\sum\limits_{r = 1}^{ M} {s_{j,r}^np_{j,r}^n} } g_{j,m}^n \le {I_b},\forall b \in \mathcal{B}}.
\end{aligned}
\end{equation}
where C1 represents the user scheduling constraint, C2 is the constraint on maximum users of the each subchannel, C3 is the constraint on the maximum associated users of each BS, C4 is the maximum power constraint on each BS, C5 is the QoS constraint on each user and  C6 is the cross-tier interference constraint on each BS.

\newcounter{TempEqCnt}
\setcounter{TempEqCnt}{\value{equation}}
\setcounter{equation}{15}

\begin{figure*}[ht]

\begin{equation}
\begin{aligned}
&   {\dfrac{{\partial {f_{X,S,P}}\left( {\mu ,\lambda ,\nu ,\tau } \right)}}{{\partial {x_{b,m}}}}=  }
&&   {\dfrac{{\log \left( {W{{\log }_2}\left( {1 + \gamma _{b,m}^n} \right)} \right)}}{{\sum\limits_{b = 1}^B {{p_{c,b}}}  + \sum\limits_{b = 1}^B {\sum\limits_{m = 1}^M {\sum\limits_{n = 1}^N {{x_{b,m}}s_{b,m}^np_{b,m}^n} } } }}
 - {\lambda _b}\left( t \right)\sum\limits_{n = 1}^N {s_{b,m}^np_{b,m}^n} \left( t \right)}\\
&&&   { - {\mu _b}\left( t \right)  + {\nu _b}\left( t \right)\sum\limits_{n = 1}^N {c_{b,m}^n}
 - {\tau _b}\left( t \right)\sum\limits_{j \in B,b \ne j} {\sum\limits_{n = 1}^N {g_{b,m}^np_{b,m}^n} \left( t \right)}}.
\end{aligned}
\end{equation}
\hrulefill
\end{figure*}
\setcounter{equation}{\value{TempEqCnt}}

\section{Machine Learning For Wireless Resource Management}

\subsection{User Association}
In this subsection, an user association scheme based on the Lagrange dual decomposition method is proposed.

The Lagrangian dual function for user association can be written as
\begin{equation}
{\rm{D}}\left( {\mu , \lambda , \nu , \tau } \right) = \mathop {\max }\limits_{X,S,P} {\rm{ }}L(\left\{ {{x_{b,m}}} \right\},\left\{ {{p_{b,m}^n}} \right\},\mu ,\lambda ,\nu ,\tau ),
\end{equation}
where $\mu {\rm{,}}\lambda {\rm{,}}\nu {\rm{,}}\tau$ are the Lagrange multipliers to decouple constraint. Then the original problem can be divided into two sub-problems by dual method,
\begin{equation}
\mathop {{\rm{min}}}\limits_{\mu ,\lambda ,\nu ,\tau } {\rm{D}}\left( {\mu ,\lambda ,\nu ,\tau } \right) = {f_{X,S,P}}\left( {\mu ,\lambda ,\nu ,\tau } \right) + {g_{K,S,P}}\left( {\mu ,\lambda ,\nu ,\tau } \right).
\end{equation}

For the sub-problem of user association the partial derivative of  ${f_{X,S,P}}\left( {\mu ,\lambda ,\nu ,\tau } \right)$ is given by (16) at the top of next page.

\setcounter{equation}{16}

The decision reference value of user association of iteration $t$  is given by
\begin{equation}
\begin{aligned}
&   {    {j_{b,m}} =  }
& & { \dfrac{{\log \left( {W{{\log }_2}\left( {1 + \gamma _{b,m}^n} \right)} \right)}}{{\sum\limits_{b = 1}^B {{p_{c,b}}}  + \sum\limits_{b = 1}^B {\sum\limits_{m = 1}^M {\sum\limits_{n = 1}^N {{x_{b,m}}s_{b,m}^np_{b,m}^n} } } }}} \\
&&& {- {\lambda _b}\left( t \right)\sum\limits_{n = 1}^N {s_{b,m}^np_{b,m}^n} \left( t \right) - {\mu _b}\left( t \right)+ {\nu _m}\left( t \right)\sum\limits_{n = 1}^N {c_{b,m}^n} }\\
&&& {- {\tau _b}\left( t \right)\sum\limits_{j \in {\cal B},b \ne j} {\sum\limits_{n = 1}^N {g_{b,m}^np_{b,m}^n} \left( t \right)}}.
\end{aligned}
\end{equation}

Then, the user association strategy can be determined as:
\begin{equation}
{x_{b,m}} = \left\{ \begin{array}{l}
1,{\rm{   }}\begin{array}{*{20}{c}}
{if}&{b = \mathop {\arg \max }\limits_m {j_{b,m}}}
\end{array}\\
0,{\rm{  }}\begin{array}{*{20}{c}}
{if}&{b \ne \mathop {\arg \max }\limits_m {j_{b,m}}}
\end{array}.
\end{array} \right.
\end{equation}

%

\subsection{Subchannel Allocation}

In this subsection, a novel co-training semi-supervised learning method for subchannel allocation is proposed.
In order to solve the channel allocation problem, we transform problem (\ref{EE}) into a loss function in the deep learning model, and get the optimal power allocation by minimizing the loss function.
The loss function can be written as
\begin{equation}
\begin{array}{l}
\mathop {\min }\limits_{\widehat S} {\left\| {\widehat S{\rm{ - }}\mathop {\arg \max }\limits_S \sum\limits_{b = 1}^B {\sum\limits_{m = 1}^M {\sum\limits_{n = 1}^N {
\dfrac{{R}}{{U}}
 } } } } \right\|^2}\\
s.t.    ~~ C2, C4-C6
\end{array}
\end{equation}
where ${\widehat S}$ means the predicted subchannel allocation strategy.


In this paper, the subchannel allocation scheme is initially generated by two-sided matching scheme as shown in Algorithm \ref{algorithm:2}. Users and subchannels are regarded as two players pursuing the maximization of their respective utility.

\begin{algorithm}[!ht]
\caption{Two-side Matching Algorithm}
\begin{algorithmic}[1]
\STATE  Initialize allocation strategy $S$;
\FOR    {$b$ in $\mathcal{B}$}
\STATE  Initialize the set $V_b(n)$ as users allocated to subchannel $n$, and set $\overline {{V_b}}$ as users of unassigned channels;
\WHILE {$\overline {{V_b}}  \ne \emptyset$ }
\FOR    {$m$ in $\mathcal{M}$}
\IF {$x_{b,m}=1$}
\STATE Select the subchannel $n^*$ with the best channel conditions.
\IF {$V_b(n^*)\ne 2$}
\STATE Let $S_{b,m}^{n^*}=1$;
\STATE Update $V_b(n^*)$ and $\overline {{V_b}}$;
\ENDIF
\IF {$V_b(n^*)= 2$}
\STATE Calculate the utility function of any two of user $m$, user $m1$,  and $m2$  who occupy subchannel $n^*$;
\STATE Select two users which maximize the EE of the subchannel $n^*$, and update $V_b(n^*)$ and $\overline {{V_b}}$;
\ENDIF
\ENDIF
\ENDFOR
\ENDWHILE
\ENDFOR
\end{algorithmic}
\label{algorithm:2}
\end{algorithm}

The proposed deep learning scheme for subchannel allocation is based on semi-supervised learning regression algorithm. In order to improve the accuracy of DNN, semi-supervised learning is used to improve the performance of NOMA heterogeneous network.
Given a labeled data set ${L_S} = \left\{ {\left( {{G_1},{S_1}} \right),\left( {{G_2},{S_2}} \right), \cdots ,\left( {{G_{\left| L \right|}},{S_{\left| L \right|}}} \right)} \right\}$ generated by Algorithm \ref{algorithm:2} and an unlabeled data set $U = \left\{ {G_1^{'},G_2^{'}, \cdots ,G_{\left| U \right|}^{'}} \right\}$, where ${G_i}$ is matrix representing the i-th group of gains, which contains all $g_{b,m}^{n}$ generated by the i-th initialization, ${S_i}$ is is the i-th group of channel allocation matrix, which contains all $s_{b,m}^{n}$, ${\left| L \right|}$ and  ${\left| U \right|}$ are the number of data samples $L_S$ and $U$, respectively.
We aim to get a learner ${H_S}:G \to S$ that can accurately predict the real label of an unlabeled input.
The input of the learner is $G$, and the output of the learner is $S$.

D. J. Miller and H. S. Uyar analyzed the feasibility and rationality of using unlabeled data to improve learners performance based on data distribution estimation theory \cite{SSL1}.
Assuming that all data follow a mixture of $L$ Gaussian distributions, then the distribution is
\begin{equation}
{H_S}\left( {\left. g \right|\theta } \right) = \sum\limits_{i = 1}^L {{\alpha _i}} {H_S}\left( {\left. g \right|{\theta _i}} \right),
\end{equation}
where $\sum\limits_{i = 1}^L {{\alpha _i}} $ is a mixing coefficient and $\theta  = \left\{ {{\theta _i}} \right\}$ is the parameter.
The random variable is a label that is determined by the selected mixture component $m_i$ and feature vector $g_i$ in the probability of $P\left( {\left. {{c_i}} \right|{g_i},{m_i}} \right)$.
According to the maximum posteriori probability hypothesis, the optimal classification formula can be written as
\begin{equation}
H\left( g \right) = \mathop {\arg \max }\limits_k \sum\nolimits_j {P\left( {\left. {{c_i} = k} \right|{g_i},{m_{i = j}}} \right)} P\left( {\left. {{m_i} = j} \right|{g_i}} \right),
\end{equation}
where $P\left( {\left. {{m_i} = j} \right|{g_i}} \right) = \dfrac{{{\alpha _j}{H_S}\left( {\left. {{g_i}} \right|{\theta _j}} \right)}}{{\sum\limits_{l = 1}^L {{\alpha _i}} {H_S}\left( {\left. {{g_i}} \right|{\theta _l}} \right)}}.$

According to the description above, the purpose of semi-supervised learning can be interpreted as using training data to estimate ${P\left( {\left. {{c_i} = k} \right|{g_i},{m_{i = j}}} \right)}$ and $P\left( {\left. {{m_i} = j} \right|{g_i}} \right)$. Obviously, the former is related to labels while the latter is not related to labels. Therefore, if there is massive unlabeled data, the generalization ability of learners is improved.
The analysis results of T. Zhang and F. J. Oles \cite{SSL2} showed that if a parameterized model can be decomposed into the form of $P\left( {\left. {x,y} \right|\theta } \right) = P\left( {\left. y \right|x,\theta } \right)P\left( {\left. x \right|\theta } \right)$, the simulated model parameters can be better supported by unlabeled data, thus improving the performance of the learning model.

Co-training semi-supervised learning algorithm is the most important semi-supervised learning method \cite{CO}. It initializes more than two learners based on clustering hypothesis or popular hypothesis. In the process of learning, the labeled data with the highest confidence after labeling is selected according to specific criteria. After labeling, the labeled data are put into each other's labeled data set of the other's learner to update the model.
Co-training method is also based on the compatibility and complementarity of multi-view data \cite{coreg}. Assuming that the data has two sufficient redundant and conditionally independent views. Sufficient means that each view contains enough information to produce the optimal learner, conditionally independent is that in given labels, each view is independent.


 \begin{algorithm}[t]
\caption{Co-training Semi-Supervised Deep Learning based Subchannel Allocation}
\hspace*{0.02in} {\bf Input:} Labeled datasets of subchannel allocation $L_S$, unlabeled dataset $U$; maximum number of iterations $T_{max}$; number of neurons in the hidden layers $n_1$, $n_2$;
\begin{algorithmic}[1]
\STATE  Initialize and train two neural network models $H_1^S$ and $H_2^S$ using datasets $L_1^S$ and $L_2^S$ that are copied from dataset $L_S$;
\STATE  Randomly select an unlabeled datapool ${U^{'}}$ of size $s$ from unlabeled dataset $U$;
\REPEAT
\FOR    {$k=1$ and 2}
\FOR    {each ${g_u} \in U^{'}$}
\STATE  ${\widehat S_u} \leftarrow H_k^S\left( {{G_u}} \right)$;
\STATE  Train network models $H_k^{S{'}}$ using datasets $L_k^S \cup \left\{ {\left( {{G_u},{{\hat S}_u}} \right)} \right\}$;
\STATE  $\delta _{{G_u}}^S = \sum\limits_{{G_i} \in L_k^S} { {{{\left( {{S_i} - H_k^S\left( {{G_i}} \right)} \right)}^2} - {{\left( {{S_i} - H{{_k^S}^{'}}\left( {{G_i}} \right)} \right)}^2}}}$;
\ENDFOR
\IF    {there exists  $\delta _{{G_u}}^S>0$}
\STATE  {$\widetilde G_k^S = \mathop {\arg \max }\limits_{{G_u} \in U_{\rm{S}}^{'}} \delta _{{G_u}}^S$; ${\widetilde S_k} = H_k^S\left( {\widetilde G_k^S} \right)$; \\
${\omega _k} = \left\{ {\left( {\widetilde G_k^S,{{\widetilde S}_k}} \right)} \right\}$;  ${U^{'}} = {U^{'}} - {\omega _k}$;}
\ELSE
\STATE {${\omega _k} =\emptyset $}
\ENDIF
\ENDFOR
\STATE  {$L_{\rm{1}}^S \leftarrow L_{\rm{1}}^S \cup {\omega _{\rm{2}}}$,
$L_{\rm{2}}^S \leftarrow L_{\rm{2}}^S \cup {\omega _{\rm{1}}}$; }
\IF    {one of $L_{\rm{1}}^S$ and $L_{\rm{2}}^S$ changes}
\STATE Update neural network models $H_1^S$ and $H_2^S$;
\STATE Replenish datapool ${U^{'}}$ to size $s$ by randomly picking;
\ENDIF
\UNTIL {Convergence or the number of iterations reaches $T_{max}$.}
\end{algorithmic}
\hspace*{0.02in} {\bf Output:} ${H^S}\left( G \right) = \frac{1}{2}\left( {H_1^S\left( G \right) + H_2^S\left( G \right)} \right)$
\label{algorithm:3}
\end{algorithm}

The co-training  semi-supervised deep learning based subchannel allocation scheme is shown in Algorithm \ref{algorithm:3}.
The key of the co-training is to define the rule for selecting high confidence labeled data. To solve this problem, the criterion defined by Algorithm \ref{algorithm:3} is that the unlabeled data with the most consistent labeled training set is the data with the highest confidence. The rules for calculating confidence are as follows:
\begin{equation}
{\Delta _u} = \frac{1}{{\left| L \right|}}\sum\limits_{{x_i} \in {L_S}} {{{\left( {{S_i} - H\left( {{G_i}} \right)} \right)}^2}}  - \frac{1}{{\left| L \right|}}\sum\limits_{{x_i} \in {L_S}} {{{\left( {{S_i} - {H^{'}}\left( {{G_i}} \right)} \right)}^2}},
\end{equation}
where $H$ represents learning model, ${g_u} \in U$, $H^{'}$ represents the retrained learning model using the original labeled data set adding labeled data generated by model $H$.

\begin{figure}[!ht]
        \centering
        \includegraphics*[width=8cm]{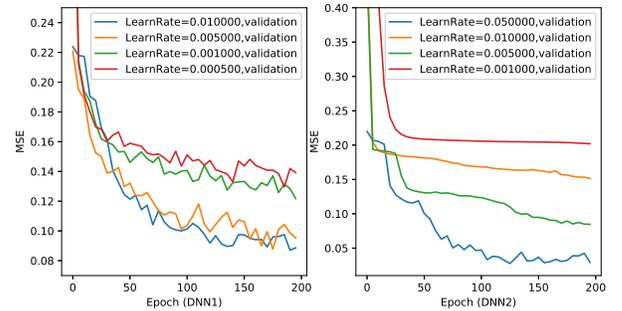}
       \caption{The mean square errors of two initial networks model DNN1 and DNN2 at different learn rate.}
        \label{fig:3}
\end{figure}

\begin{figure}[!ht]
        \centering
        \includegraphics*[width=8cm]{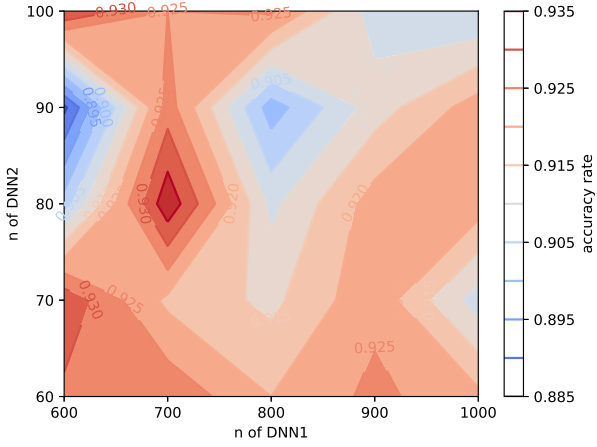}
       \caption{The accuracy rate of co-training semi-supervised learning based subchannel allocation with different numbers of neurons per hidden layer of two learners.}
        \label{fig:4}
\end{figure}

According to Zhou's research results \cite{coreg}, in order to start a co-training process with good effects, the two initialized learners must have relatively large differences. In extreme cases, if two learners are very close, the added labeled data will be very similar, and eventually it will evolve into a single learner, that is, self-training algorithm.
When designing learners of Algorithm \ref{algorithm:3}, we ensure the largest difference between learner $H_1^S$ and learner $H_2^S$.
The differences of the models are mainly obtained by using different hidden layers, different network parameters or different number of nodes on the hidden layer.

Two network models are designed as 3 hidden layer network and 4 hidden layer network, the dropout coefficients are 0.8 and 1, batch-size are 200 and 500, respectively. We run the initial learners with different learning rates and choose the better learning rates of 0.01 and 0.05 respectively, as shown in Fig. \ref{fig:3}. The Algorithm \ref{algorithm:3} runs under different number of hidden layer neurons from 600 to 1000 and from 60 to 10 in order to choose the optimal parameters, and the accuracy of the learner is evaluated as shown in the Fig. \ref{fig:4}. By comparing different accuracy rates,  the  neurons number of hidden layer in the two networks is $n_1=700$ and $n_2=80$, respectively.

\subsection{Power Allocation}

In this subsection, a power optimization scheme based on DNN is proposed to maximize the EE.
DNN scheme has been proved to have high approximation rate and computing performance for power control problem in \cite{haoransun2018}.
In this scheme, the complex iteration process is replaced by the non-linear mapping.
Its output operation is realized only by simple matrix multiplication and addition, which guarantees the real-time performance of the network.
DNN is a multi-layer network model, which works mainly through the connection between neurons. Specifically,  denote $\xi$ as the activation function, denote $n_j$ as the number of neural of layer $j$, denote $w_i^j$ as the weight coefficient and $b^j$ as the bias of layer $j$. The output of the ${i^{{\rm{th}}}}$ neural of layer $j$ is given by
\begin{equation}
y_i^j = \xi \left( {\sum\limits_{i = 1}^{{n_j}} {w_i^{j - 1}y_i^{j - 1} + {b^j}} } \right).
\end{equation}

Consider the loss function of power optimization problem, problem (\ref{EE}) can be rewritten as
\begin{equation}
\begin{array}{l}
\mathop {\min }\limits_{\widehat P} {\left\| {\widehat P{\rm{ - }}\mathop {\arg \max }\limits_P \sum\limits_{b = 1}^B {\sum\limits_{m = 1}^M {\sum\limits_{n = 1}^N {
\dfrac{{R}}{{U}}
} } } } \right\|^2}    \\
s.t. ~~ C4-C6
\end{array}
\end{equation}
where ${\widehat P}$ shows predicted matrix of power optimization.
We translate the maximization problem of non-deterministic polynomials into the problem of minimizing the error between the predicted value and the sample value, which is also directly expressed as the problem of the mean square error between the predicted value and the sample that represents the accuracy of the network model.


An iterative gradient algorithm is used to generate training data.
Let ${p_{m,b}} = \sum\limits_{n = 1}^N {s_{b,m}^np_{b,m}^n} $, and the power optimization function is given by:
\begin{equation}
\begin{aligned}
&   {f({p_{b,m}}) =} \\
&  {\sum\limits_{b = 1}^B {\sum\limits_{m = 1}^M {\sum\limits_{n = 1}^N {{x_{b,m}}\dfrac{{\ln \left( {{{\log }_2}\left( {1 + \gamma _{b,m}^n} \right)} \right) - \ln \left( {{K_b}} \right)}}{{\sum\limits_{b = 1}^B {{p_{c,b}}}+ \sum\limits_{b = 1}^B {\sum\limits_{m = 1}^M {\sum\limits_{n = 1}^N {{x_{b,m}}s_{b,m}^np_{b,m}^n} } } }}} } }}\\
& {+\sum\limits_{b = 1}^B {{\lambda _b}\left( {{p_{\max }} - \sum\limits_{m = 1}^M {\sum\limits_{n = 1}^N {{x_{b,m}}s_{b,m}^np_{b,m}^n} } } \right)}} .
\end{aligned}
\end{equation}

In this case, the power ${p_{b,m}}$ is updated shown below:
\begin{equation}
{p_{b,m}}\left( {t + 1} \right) = {p_{b,m}}\left( t \right) + \delta \left( t \right) \cdot \Delta {p_{b,m}},
\end{equation}
where ${\delta}\left( t \right)$ is the step and $\Delta {p_{b,m}}$ is shown as
\begin{equation}
\Delta {p_{b,m}}=\left| {\dfrac{{\partial f}}{{\partial {p_{b,m}}}}} \right|{\rm{ /}}\left| {\dfrac{{{\partial ^2}f}}{{\partial {p_{b,m}}^2}}} \right|.
\end{equation}

\begin{algorithm}[t]
\caption{Deep Neural Networks Based Power Optimization Scheme}
\hspace*{0.02in} {\bf Data Generation:}
\begin{algorithmic}[1]
\FOR    {$n = 1$ to the size of data set}
\REPEAT
\FOR    {$b$ in $\mathcal{B}$}
\FOR    {$m$ in $\mathcal{M}$}
\STATE  (1)	Calculate $\Delta {p_{b,m}}$=$\dfrac{{\partial f}}{{\partial {p_{b,m}}}}$ / $\dfrac{{{\partial ^2}f}}{{\partial {p_{b,m}}^2}}$;
\STATE  (2)	${p_{b,m}}\left( {t + 1} \right) = {p_{b,m}}\left( t \right) + {\delta }\left( t \right)\Delta {p_{b,m}}$;
\ENDFOR
\ENDFOR
\UNTIL {Convergence or $t = {T_{\max }}$};
\ENDFOR
\end{algorithmic}
\hspace*{0.02in} {\bf Training Stage:}
\begin{algorithmic}[1]
\STATE  Initialize 3 layers DNN structure with $n_j$ neurons in each layer, the weight $w$ and bias $b$.
\FOR    {$m=1$ to training-epochs}
\FOR    {$n = 1$ to batch-size}
\STATE  Update the weight $w$ and bias $b$;

The activation function: RELU;

The optimization algorithm: Adam algorithm.

\ENDFOR
\ENDFOR
\end{algorithmic}
\hspace*{0.02in} {\bf Testing Stage:}
\begin{algorithmic}[1]
\STATE  Generate the testing dataset.
\STATE  Pass testing dataset through the trained power optimization model.
\STATE  Evaluate the performance through DNN model.
\end{algorithmic}
\label{algorithm:4}
\end{algorithm}

In training stage of DNN based power optimization scheme, we update the weight $w$ and bias $b$ to minimize lost function. The process of DNN based power optimization scheme is shown as Algorithm \ref{algorithm:4}.
The rectified linear unit (RELU) is used as the activation function and the optimization algorithm we use is the Adam algorithm.
The related neural network structure and simulation environment are discussed in the next section.

\subsection{Complexity Performance}
In this subsection, the asymptotic time complexity of proposed radio resource management is discussed.

\begin{itemize}
  \item The complexity of  user association algorithm.
\end{itemize}

The proposed scheme is based on the Lagrange dual decomposition method.
The complexity performance of the Lagrange dual decomposition method for user association is discussed in \cite{JSAC2017}.
Thus, user association needs $O\left( M B + 3B +M \right)$ operations at each iteration.

\begin{itemize}
  \item The complexity of subchannel allocation algorithm.
\end{itemize}

The proposed deep learning scheme for subchannel allocation is based on semi-supervised learning regression algorithm, which contains two learner $H_1^S$ and $H_2^S$ of different number of hidden layers and nodes per layer.
The time complexity of DNN can be represented by floating-point operations (FLOPs). For each layer of neural network, the number of FLOPs can be expressed as
\begin{equation}
FLOPs = 2{I_i}{O_i},
\end{equation}
where $I_i$ is the input dimension of the $i$th layer and $O_i$ is the output dimension of the $i$th layer.
Therefore, for our semi-supervised learning scheme, the number of FLOPs is:
\begin{equation}
\begin{array}{l}
FLOPs = 2\left( {\sum\limits_{i = 1}^3 {{I_i}{O_i}}  + \sum\limits_{i = 1}^4 {{I_i}{O_i}} } \right)\\
 = 2B\left( {\underbrace {M/B(N + 1){n_1} + n_1^2}_{{\rm{DNN model1}}} + \underbrace {M/B(N + 1){n_2} + 2n_2^2}_{{\rm{DNN model2}}}} \right).
\end{array}
\end{equation}
While the time complexity of two-side matching algorithm would be $O(B N!{2^{M/B}})$ discussed in \cite{Noma}.
So the complexity comparison between learning-based subchannel allocation and matching subchannel allocation is as follows:
\begin{equation}
\begin{array}{l}
O(2B\left( {M/B(N + 1)({n_1} + {n_2}) + n_1^2 + 2n_2^2} \right)) \\ < O(BN!{2^{M/B}}).
\end{array}
\end{equation}
Therefore, the proposed learning model for subchannel allocation has a lower complexity than matching algorithm.

\begin{itemize}
  \item The complexity of power allocation algorithm.
\end{itemize}

For our learning scheme for power allocation, the number of FLOPs is: \begin{equation}
FLOPs = 2\left( {\sum\limits_{i = 1}^3 {{I_i}{O_i}} } \right) = 2B\left( {M/B(N + 1){n_j} + n_j^2} \right).
\end{equation}
The complexity performance of the iterative gradient algorithm for power allocation is discussed in \cite{JSAC2017}.
Thus, data generation of Algorithm 4 needs $O\left( M B + B \right)$  operations at each iteration.
We suppose that the iterative gradient algorithm requires $t_p$ iterations to converge.

So the complexity comparison between learning-based power allocation and iterative gradient algorithm is as follows:
\begin{equation}
O\left( {2B\left( {M/B(N + 1){n_j} + n_j^2} \right)} \right) < O\left( {t_p \left( {BM + B} \right)} \right).
\end{equation}
Therefore, the proposed learning model for power allocation has a lower complexity than the iterative gradient algorithm.

\section{Simulation Results And Discussion}

In this section, we simulated the performance of the proposed deep learning-based resource optimization algorithm.
In our simulation, a NOMA-based mmWave network with all users distributed within the coverage of the macrocell is studied.
The power of AWGN is ${\sigma ^2}=-134$ dBm; system bandwidth is 1200 MHz; the radius of the macrocell is 100 m; up to 2 users can be occupied on each subchannel; the maximum power of the macrocell is 9.5 dBm, and the maximum power of the small cells is 4.7 dBm.
The parameters for deep learning model are as shown in Table I.

\begin{table}[htbp]
\centering
 \caption{\label{tab:test}The parameters for deep learning model}
\begin{tabular}{ l | c | c }
\toprule
\multirow{2}*{Symbols}&
		\multicolumn{2}{c}{Values} \\
		\cline {2-3}
		        & Algorithm 2 & Algorithm 3  \\
 \midrule  The number of training sets  & 5000 & 5000 \\
The number of testing sets  & 1000 & 1000 \\
  The number of layers  & 3 / 4 & 3 \\
 The number of neurons per layer & 700 / 80 & 800 \\
  Learn rate & 0.01 / 0.05 & 0.01\\
Batch size & 200 / 500 & 200 \\
Epochs   & 100 & 100 \\
Optimizer &  RMSProp &  Adam \\
\bottomrule  \end{tabular}
 \end{table}
And the simulation environment of the proposed scheme is as follows: Python 3.6 with TensorFlow 1.3.0 with NVIDIA GeForce GTX 1050.

\begin{figure}[h]
        \centering
        \includegraphics*[width=8cm]{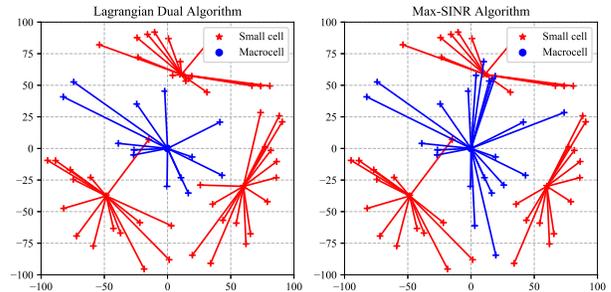}\\
       \caption{Relationship between user and base station by  Lagrange dual decomposition and Max-SINR algorithm.}
        \label{fig:1}
\end{figure}
Fig. \ref{fig:1} shows the user association strategy obtained by Lagrange dual decomposition and Max-SINR algorithm. The network environment includes one macrocell BS, three small BSs and 60 users.
The main purpose of user association is to solve the problem of load balancing and blind area coverage.
By comparing users connected with different types of base stations, it can be seen that under max-SINR algorithm, most users are assigned to macrocell, indicating that the load balancing effect of this algorithm is the worst. Under the Lagrangian Dual method, the number of users connected to macrocell is reduced.
Compared to max-SINR algorithm, Lagrange dual decomposition method based user association algorithm embodies better performance of load balancing.
%
%

%
%

\begin{figure}[h]
        \centering
        \includegraphics*[width=8cm]{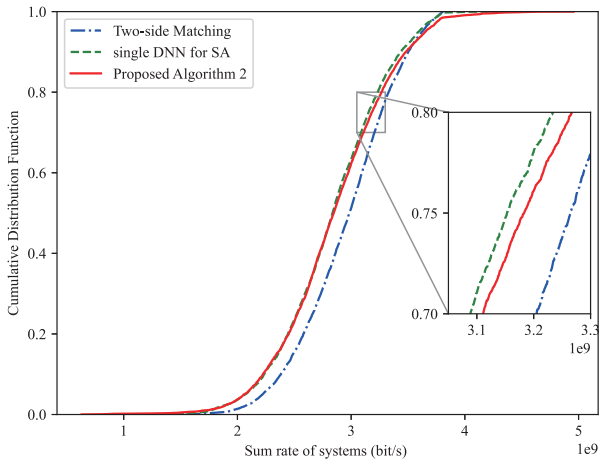}
       \caption{The CDF that describes the sum rate of system achieved by different approach:       1) two-side matching algorithm;       2) single DNN algorithm for subchannel allocation;       3) semi-supervised learning for subchannel allocation. }
        \label{fig:88}
\end{figure}

As shown in the Fig. \ref{fig:88}, the sum rate of the system is evaluated for different methods.
Each line represents the results obtained from 5,000 random data test points.
The distributions of sum rate under different algorithms are consistent approximately, ranging from $2.0 \times {10^{9}}$ to $3.6 \times {10^{9}}$.
It is obviously that the results of deep learning schemes are close to the result of two-side matching algorithm.
Furthermore, it is also obviously that the performance of semi-supervised learning scheme is much better than that of single DNN models, which proves that semi-supervised learning can address the problem of subchannel allocation well.

\begin{figure}[h]
        \centering
        \includegraphics*[width=8cm]{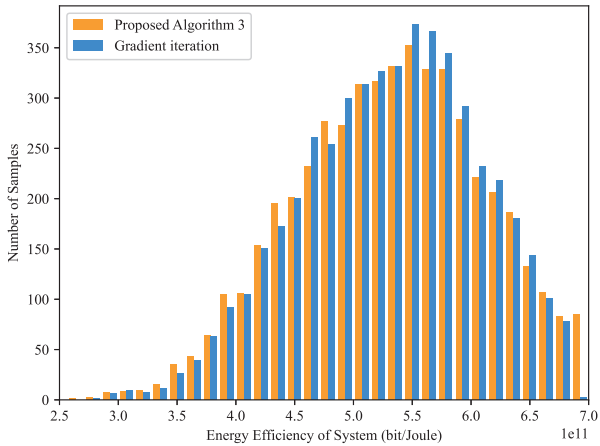}
       \caption{Distributions of the EE of system by DNN approach for power allocation and gradient iteration algorithm.}
        \label{fig:5}
\end{figure}

Fig. \ref{fig:5} shows the distribution of the EE of system by DNN approach and the gradient iteration algorithm over the whole test dataset. DNN approach gives a good behavior of approximation (about 98\%) of the EE generated by gradient iteration algorithm.

\begin{figure}[h]
        \centering
        \includegraphics*[width=8cm]{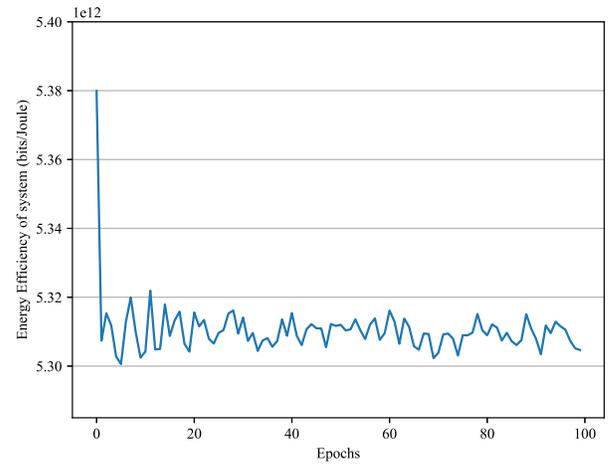}
       \caption{The EE of systems versus number of training sequences.}
        \label{fig:78}
\end{figure}
Fig. \ref{fig:78} shows the EE of systems versus number of training sequences.
It is obviously that the results of DNN schemes converges to $5.31 \times 10^{12}$ bits/joule with the changes of epochs, during the training process.

\begin{figure}[h]
        \centering
        \includegraphics*[width=8cm]{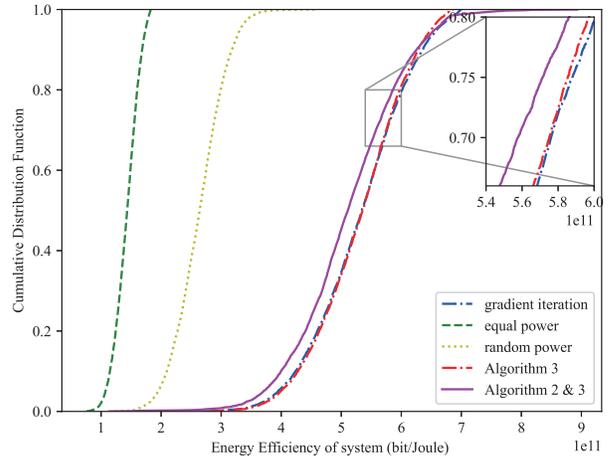}
       \caption{The CDF that describes the EE of system achieved by different approach:       1) gradient iteration algorithm;       2) equal power allocation;       3) random power allocation;       4) DNN model for power optimization;       5) semi-supervised learning for subchannel allocation \& DNN model for power optimization. }
        \label{fig:8}
\end{figure}
As shown in the Fig. \ref{fig:8}, we evaluate the EE of system achieved by different schemes.
Each line represents the results obtained from 5,000 random data test points.
The range of EE of the system under deep learning schemes is mostly distributed between $0.4 \times {10^{12}}$ to $0.6 \times {10^{12}}$.
It is obviously that the results of DNN schemes are close to the results of gradient iteration algorithm, significantly superior to the other equal power scheme and random power scheme.
Furthermore, Fig. \ref{fig:8} shows that semi-supervised learning for subchannel allocation and DNN model for power optimization perform well in EE optimization, and also proves that our proposed schemes can address the resource optimization problem well.

\section{Conclusions}
In this paper, a framework based on deep learning has been designed for dealing with user association, subchannel allocation, and power control in NOMA mmWave networks.
user association is solved by the Lagrange dual decomposition, the subchannel allocation is solved by semi-supervised learning and power allocation is solved by DNN scheme.
In the proposed scheme, sample data are generated by iterative algorithm, and ML schemes are adopted in the decision-making stage.
The proposed scheme improves the EE of the system, and reduces the dependence on the reference algorithm by using unlabeled data.
Simulation results verify that in the NOMA-based mmWave network, the effectiveness of the proposed deep learning method in user association, subchannel allocation and power control.

\section*{Acknowledgment}
This work is supported by the National Key R\&D Program of China (2019YFB1803304), National Natural Science Foundation of China (61822104, 61771044),  Beijing Natural Science Foundation (L172025, L172049), and the Fundamental Research Funds for the Central Universities(FRF-TP-19-002C1, RC1631), Beijing Top Discipline for Artificial Intelligent Science and Engineering, University of Science and Technology Beijing.

\end{document}